\documentclass[aps,prl,twocolumn,superscriptaddress,floatfix,nofootinbib]{revtex4}
\usepackage{amsfonts,hyperref}
\usepackage{amsmath}
\usepackage{amssymb}

\usepackage{graphicx}
\usepackage{amsmath}
\usepackage{amssymb}

\begin{document}

\title{
Comment on "Topological quantum phase transitions of attractive spinless fermions in a honeycomb lattice" by Poletti D. et~al.
}

\author{Philippe Corboz}
\affiliation{Theoretische Physik, ETH Zurich, 8093 Zurich, Switzerland}
\affiliation{Institute of Theoretical Physics, \'Ecole Polytechnique F\'ed\'erale de Lausanne (EPFL), CH-1015 Lausanne, Switzerland}

\author{Sylvain Capponi}
\affiliation{Laboratoire de Physique Th\'eorique (IRSAMC), CNRS and Universit\'e de Toulouse, F-31062 Toulouse, France}

\author{Andreas M. L\"auchli}
\affiliation{Institut f\"ur Theoretische Physik, Universit\"at Innsbruck, A-6020 Innsbruck, Austria}
\affiliation{Max-Planck-Institut f\"{u}r Physik komplexer Systeme, N\"{o}thnitzer Stra{\ss}e 38, D-01187 Dresden, Germany}

\author{Bela Bauer}
\affiliation{Station Q, Microsoft Research, Santa Barbara, CA 93106}

\author{Roman Or\'us}
\affiliation{Max-Planck-Institut f\"ur Quantenoptik, Hans-Kopfermann-Stra{\ss}e 1, 85748 Garching, Germany}

\date{\today}

\begin{abstract}
\end{abstract}


\maketitle
In a recent letter \cite{poletti2011} a model of attractive spinless fermions on the honeycomb lattice at half filling has been studied by mean-field theory, where distinct homogenous phases at rather large attraction strength $V>3.36$, separated by (topological) phase transitions, have been predicted. In this comment we argue that without  additional interactions the ground states in these phases are not stable against phase separation. We determine the onset of phase separation at half filling $V_{ps}\approx 1.7$ by means of infinite projected entangled-pair states (iPEPS)~\cite{verstraete2004,jordan2008,corboz2010} and exact diagonalization (ED).

The Hamiltonian of the model reads
\begin{equation}
\label{eq:H}
\hat H = -t \sum_{\langle i,j \rangle} \hat c^\dagger_i c_j + h.c. - V\sum_{\langle i,j \rangle} (\hat n_i -\frac{1}{2}) (\hat n_j - \frac{1}{2}) - \mu\sum_{i} \hat n_i 
\end{equation}
with $t=1$ the hopping amplitude, and $V>0$ the attraction strength. We work in the grand-canonical ensemble, i.e. we use a chemical potential $\mu$ to control the particle density $n(\mu)$ in the system. Setting $\mu=0$ corresponds to a half-filled state, $n=0.5$, if the state at half filling is stable towards phase separation. 

Intuitively, at half filling, if the attraction $V$ is much stronger than the hopping $t$, the fermions can minimize their energy by clustering, leading to phase separation, where half of the system is empty and the other half is occupied by the fermions. In the grand-canonical ensemble, such an instability can be identified as  a discontinuity (a jump) in the particle density $n(\mu)$ at~$\mu=0$.

\begin{figure}[htb]
  \includegraphics[width=8cm]{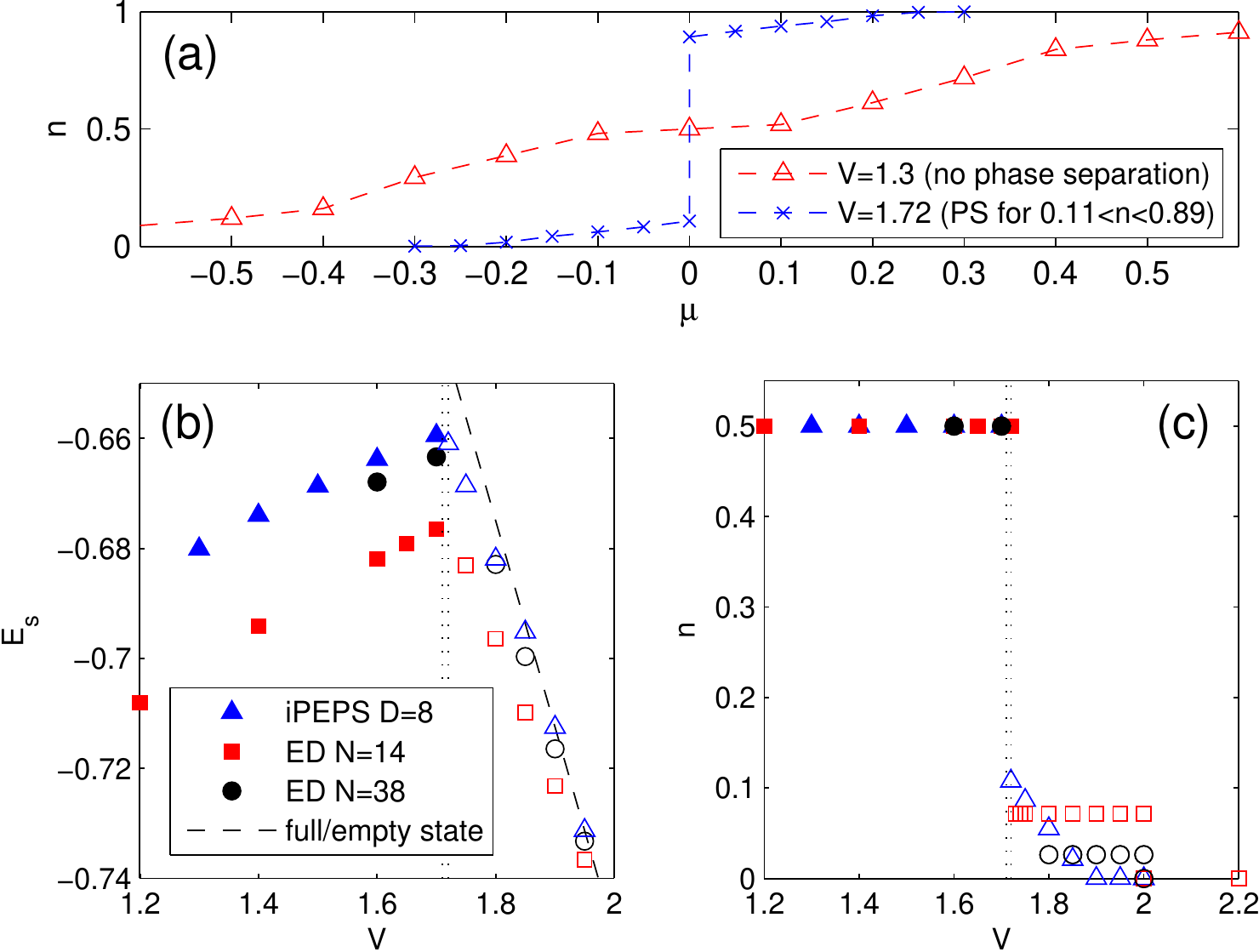}
\caption{
(a) Particle density $n$ as a function of the chemical potential $\mu$ obtained with iPEPS with a bond dimension $D=8$. For $V=1.3$ the density increases continuously with increasing chemical potential. For $V=1.72$ the particle density exhibits a jump at $\mu=0$ between the two densities $n_1\approx0.11$ and $n_2 \approx 0.89$, which indicates an unstable region (phase separation) between these two densities. 
(b)~Energy as a function of interaction strength $V$ for $\mu=0$ obtained with iPEPS and ED. The full symbols for $V<V_{ps}\approx 1.71$ correspond to stable solutions at half filling, whereas open symbols for $V>V_{ps}$ correspond to states away from half filling with a density of either $n_1(V)$ or $n_2(V) = 1-n_1(V)$. The dashed line corresponds to the energy of a completely filled or empty state. (c)~Density as a function of $V$ for $\mu=0$ obtained with iPEPS and ED. For each state with density $n_1(V)<0.5$ there is a degenerate state with a density $n_2(V)=1-n_1(V)$.
}
\label{fig:iPEPS}
\end{figure}

Figure~\ref{fig:iPEPS}(a) summarizes our numerical results obtained with ED on finite systems and with iPEPS, a tensor network ansatz to simulate the model directly in the thermodynamic limit. The accuracy of the iPEPS can be systematically controlled by the so-called bond dimension~$D$. Details on the method can be found in refs.~\cite{corboz2010,corboz2011}.

Figure~\ref{fig:iPEPS}(a) shows an example of a jump in $n(\mu)$ for $V=1.72$ between the two densities $n_1\approx0.11$ and $n_2 \approx 0.89$ obtained with iPEPS. For densities in between these two values there is no stable homogenous solution, because it is energetically favorable for the system to split into two regions, one with density $n_1$ and the other one with $n_2$. Since iPEPS is an ansatz for a homogeneous phase, we either obtain a state with density $n_1$ or a state with density $n_2$ for $\mu=0$, if there is no homogenous solution at half filling. 
For very large attraction, $V \gtrsim 1.9$, the system splits into a completely empty and a completely filled region, i.e. $n_1=0$, $n_2=1$. The dependence of $n_1$ as a function of $V$ is shown in  fig.~\ref{fig:iPEPS}(c).

The full symbols in fig.~\ref{fig:iPEPS}(b) for $V<1.71$ show the iPEPS energy of stable solutions at half filling for $\mu=0$, whereas the open symbols for $V>1.71$ correspond to states away from half-filling, where the state at half-filling is unstable. The value for the onset of phase separation $V_{ps}\approx 1.71$ depends only weakly on the bond dimension.

A similar result is found with ED, where we considered different system sizes up to $N=38$ lattice sites. Phase separation can already be seen for small systems, e.g. for $N=14$ as shown in fig.~\ref{fig:iPEPS}(b-c). The value $V_{ps}\approx 1.7$ depends only weakly on the system size.

In conclusion, we obtained consistent results with iPEPS and ED which clearly show that for attractions stronger than $V\approx 1.7$ the half-filled state is not stable, but that the system phase separates into a low-density and a high-density region.  For attractions stronger than $V\approx 1.9$ the system phase separates into a completely filled, and a completely empty region. This suggests that the homogenous phases for $V>3.36$ found in ref.~\cite{poletti2011} are not stable ground state solutions of the Hamiltonian~\eqref{eq:H}. We note, however, that these phases may possibly be stabilized by including longer-ranged (repulsive) interactions in the Hamiltonian.

\acknowledgements We thank D. Poletti, C. Miniatura, and B. Gr\'emaud
for constructive discussions. The simulations were performed on the
Brutus cluster at ETH Zurich, on the PKS-AIMS cluster at the MPG RZ
Garching, and CALMIP.


\bibliographystyle{apsrev4-1}
\bibliography{tV_honeycomb}

\end{document}